\documentclass[draft,prd,aps,showpacs]{revtex4}
\usepackage{amsmath,amssymb}
\input psfig.sty

\def\bea{\begin{eqnarray}}
\def\eea{\end{eqnarray}}
\def\beq{\begin{equation}}
\def\eeq{\end{equation}}

\def\esp{\hspace{0.3cm}}

\begin{document}

\title{Light pseudoscalar mesons in a nonlocal SU(3) chiral quark model}

\author{A. Scarpettini$^a$, D. G\'omez Dumm$^b$$^\dagger$ and Norberto
N. Scoccola$^{a,c}$\footnote[2]{Fellow of CONICET, Argentina.}}

\affiliation{
$^a$ Physics Department, Comisi\'on Nacional de Energ\'{\i}a At\'omica, \\
 Av.Libertador 8250, (1429) Buenos Aires, Argentina.\\
$^b$ IFLP, Depto.\ de F\'{\i}sica, Universidad Nacional de La Plata,
     C.C. 67, (1900) La Plata, Argentina.\\
$^c$ Universidad Favaloro, Sol{\'\i}s 453, (1078) Buenos Aires,
Argentina.}


\begin{abstract}
We study the properties of the light pseudoscalar mesons in a three flavor
chiral quark model with nonlocal separable interactions. We concentrate on
the evaluation of meson masses and decay constants, considering both the
cases of Gaussian and Lorentzian nonlocal regulators. The results are
found to be in quite good agreement with the empirical values, in
particular in the case of the ratio $f_K/f_\pi$ and the anomalous decay
$\pi^0\to\gamma\gamma$. In addition, the model leads to a reasonable
description of the observed phenomenology in the $\eta-\eta'$ sector,
even though it implies the existence of two significantly different
state mixing angles.
\end{abstract}


\maketitle
\section{Introduction}

The properties of the light pseudoscalar mesons (i.e.\ the pions,
kaons and etas) provide a suitable ground for the study of the
basic nonperturbative features of Quantum Chromodynamics (QCD). As
well known, the QCD Lagrangian shows an approximate $U(3)_L
\otimes U(3)_R$ chiral symmetry, which is spontaneously broken
down to $U(3)_V$ in the low momentum, nonperturbative regime. The
fact that, instead of nine, only eight pseudoscalar
quasi-Goldstone bosons are observed in Nature is usually explained
in terms of the so-called $U(1)_A$ anomaly. This anomaly is again
related to nonperturbative aspects of QCD, and it is believed to
be mainly responsible for the rather large $\eta'$ mass.
Unfortunately, so far it has not been possible to obtain detailed
information about the properties of the light pseudoscalar mesons
directly from QCD, and most of the present theoretical work on the
subject relies in low energy effective theories. Among them the
Nambu-Jona-Lasinio (NJL) model~\cite{NJL61} and its three flavor
extensions~\cite{KH88,BJM87,TTKK90,KLVW90} are some of the most
popular ones. In the NJL model the quark fields interact via local
effective vertices which are subject to chiral symmetry. If such
interaction is strong enough chiral symmetry is spontaneously
broken and pseudoscalar Goldstone bosons appear~\cite{reports}. As
an improvement of the local NJL scheme, some covariant nonlocal
extensions have been studied in the last few years~\cite{Rip97}.
Nonlocality arises naturally in the context of several quite
successful approaches to low-energy quark dynamics as, for
example, the instanton liquid model~\cite{SS98} and the
Schwinger-Dyson resummation techniques~\cite{RW94}. It has been
also argued that nonlocal covariant extensions of the NJL model
have several advantages over the local scheme. Indeed, nonlocal
interactions regularize the model in such a way that anomalies are
preserved~\cite{AS99} and charges properly quantized, the
effective interaction is finite to all orders in the loop
expansion and there is not need to introduce extra cut-offs, soft
regulators such as Gaussian functions lead to small
next-to-leading order corrections~\cite{Rip00}, etc. In addition,
it has been shown~\cite{BB95,PB98} that a proper choice of the nonlocal
regulator and the model parameters can lead to some form of quark
confinement, in the sense that the effective quark propagator has
no poles at real energies.

Until now, most of the research work on nonlocal chiral models has been
restricted to the flavor $SU(2)$ sector including applications to the
baryonic sector~\cite{BGR02,RWB03} and to the study of the phase
transitions at finite temperature and densities~\cite{GDS01}. The aim of
the present paper is to extend this type of models so as to include
strange degrees of freedom, and to analyze the predictions for the masses
and decay constants for the pions, kaons and the $\eta-\eta'$ system.

This article is organized as follows. In Sec.\ II we present the
general formalism and derive the expressions needed to evaluate
the different meson properties. The numerical results for some
specific nonlocal regulators together with the corresponding
discussions are given in Sec.\ III, while in Sec.\ IV we present
our conclusions. Finally, we include an Appendix with some details
concerning the evaluation of quark loop integrals.

\section{The formalism}

\subsection{Effective action}

We start by the Euclidean quark effective action
\begin{eqnarray}
S_E &=& \int d^4x \ \left\{ \bar \psi (x) \left[ -i \gamma_\mu
\partial_\mu + \hat m_c \right] \psi(x) - \frac{G}{2} \left[
j_a^S(x) \ j_a^S(x) + j_a^P(x) \ j_a^P(x) \right] \right.
\nonumber \\ & & \qquad \qquad \left. - \frac{H}{4} A_{abc} \left[
j_a^S(x) j_b^S(x) j_c^S(x) - 3\ j_a^S(x) j_b^P(x) j_c^P(x) \right]
\right\}\;,
\label{se}
\end{eqnarray}
where $\psi$ is a chiral $U(3)$ vector that includes the light
quark fields, $\psi \equiv (u\; d\; s)^T$, while $\hat m_c = {\rm
diag}(m_u, m_d, m_s)$ is the current quark mass matrix. We will
work from now on in the isospin symmetry limit, in which $m_u =
m_d$. The currents $j_a^{S,P}(x)$ are given by
\begin{eqnarray}
j_a^S (x) &=& \int d^4 y\ d^4 z \ r(y-x) \ r(x-z) \  \bar \psi(y)
\ \lambda_a \ \psi(z)\, , \\ j_a^P (x) &=& \int d^4 y\ d^4 z \
r(y-x) \ r(x-z) \  \bar \psi(y) \  i \gamma_5 \ \lambda_a \
\psi(z)\, ,
\end{eqnarray}
where the regulator $r(x-y)$ is local in momentum space, namely
\begin{equation}
r(x-y) = \int \frac{d^4p}{(2\pi)^4} \ e^{-i(x-y) p} \ r(p) \;,
\end{equation}
and the matrices $\lambda_a$, with $a=0,..,8$, are the usual eight
Gell-Mann $3\times 3$ matrices ---generators of $SU(3)$--- plus
$\lambda_0=\sqrt{2/3}\;\openone_{3\times 3}$. Finally, the
constants $A_{abc}$ are defined by
\begin{equation}
A_{abc} = \frac{1}{3!} \ \epsilon_{ijk} \ \epsilon_{mnl} \
\left(\lambda_a\right)_{im} \left(\lambda_b\right)_{jn}
\left(\lambda_c\right)_{kl}\;.
\end{equation}

The corresponding partition function $Z = \int {\cal D} \bar \psi\,
{\cal D} \psi \, \exp[- S_E]$ can be bosonized in the usual way
introducing the scalar and pseudoscalar meson fields $\sigma_a(x)$
and $\pi_a(x)$ respectively, together with auxiliary fields
$S_a(x)$ and $P_a(x)$. Integrating out the quark fields we get
\begin{eqnarray}
Z & = & \int {\cal D} \sigma_a\, {\cal D} \pi_a \ \det A \int
{\cal D} S_a\, {\cal D} P_a \ \exp\left[ \int d^4 x\; (\sigma_a
S_a + \pi_a  P_a) \right] \times \nonumber \\ & & \exp\left\{ \int
d^4 x \left[ \frac{G}{2} \left( S_a S_a + P_a P_a \right) +
\frac{H}{4} A_{abc} \left( S_a S_b S_c - 3 S_a P_b P_c \right)
\right] \right\} \; ,
\end{eqnarray}
where the operator $A$ reads, in momentum space,
\begin{eqnarray}
A(p,p') = (\,-\rlap/p + \hat m_c)\,(2\pi)^4 \,\delta^{(4)}(p-p')
 + r(p)\, \left[ \sigma_a(p-p')+ i \gamma_5 \ \pi_a(p-p')
\right]   \lambda_a \ r(p') \;.
\end{eqnarray}

To perform the integration over the fields $S_a$ and $P_a$ we use
the Stationary Phase Approximation (SPA). This means to replace
the integral over $S_a$ and $P_a$ by the integrand evaluated at
its minimizing values $\tilde S_a(\sigma_b(x),\pi_c(x))$ and
$\tilde P_a(\sigma_b(x),\pi_c(x))$. The latter are required to
satisfy
\begin{eqnarray}
& & \sigma_a + G \tilde S_a + \frac{3 H}{4} \ A_{abc} \left[
\tilde S_b \ \tilde S_c -  \tilde P_b \ \tilde P_c \right] = 0\;,
\nonumber \\ & & \pi_a + G \tilde P_a - \frac{3 H}{2} \ A_{abc} \
\tilde S_b \ \tilde P_c = 0\;. \label{spa}
\end{eqnarray}
Thus, within the SPA the bosonized effective action reads
\begin{equation}
S_E = - \,\ln\, {\rm det}\, A - \int d^4 x \ \bigg[ \sigma_a
\tilde S_a + \pi_a  \tilde P_a + \frac{G}{2} \left( \tilde S_a
\tilde S_a + \tilde P_a \tilde P_a \right) +
    \frac{H}{4} A_{abc}
    \left( \tilde S_a \tilde S_b \tilde S_c - 3\,\tilde S_a \tilde P_b
\tilde P_c \right) \bigg]\;. \label{bosonact}
\end{equation}

At this stage we assume that the $\sigma_a$ fields can have
nontrivial translational invariant mean field values
$\bar\sigma_a$ while the pseudoscalar field cannot. Thus, we write
\begin{eqnarray}
\sigma_a(x) &=& \bar \sigma_a + \delta \sigma_a(x)\;, \nonumber \\
\pi_a(x) &=& \delta \pi_a (x)\;. \label{mfa}
\end{eqnarray}
Note that due to charge conservation only $\bar \sigma_{a=0,3,8}$
can be different from zero. Moreover, $\sigma_3$ also vanishes in
the isospin limit. After replacing Eqs.~(\ref{mfa}) in the
bosonized effective action (\ref{bosonact}) and expanding up to
second order in the fluctuations $\delta \sigma_a(x)$ and $\delta
\pi_a (x)$ we get
\begin{eqnarray}
S_E = S_E^{MFA} + S_E^{quad} + \ \dots \label{act}
\end{eqnarray}
Here the mean field action reads
\begin{equation}
\frac{S_E^{MFA}}{V^{(4)}} = -\, 2\, N_c  \int \frac{d^4
p}{(2\pi)^4} \ {\rm Tr}\, \ln \left[ p^2\,\openone_{3\times 3} +
\Sigma^2(p) \right] - \frac{1}{2} \left[ \sum_i \left( \bar
\sigma_i \bar S_i  + \frac{G}{2} \bar S_i \bar S_i \right) +
\frac{H}{2} \bar S_u \bar S_d \bar S_s \right]\;,
\label{semfa}
\end{equation}
where for convenience we have changed to a new basis in which $S_i$, with
$i=u,d,s$ (or equivalently $i=1,2,3$) are given by
\begin{eqnarray}
S_u & = & \sqrt{\frac23} \, S_0 + S_3 + \frac1{\sqrt{3}} \, S_8
\nonumber \; , \quad S_d  =  \sqrt{\frac23} \, S_0 - S_3 +
\frac1{\sqrt{3}} \, S_8 \nonumber \; , \quad S_s  =
\sqrt{\frac23} \, S_0 + \frac2{\sqrt{3}} \, S_8\;,
\end{eqnarray}
and similar definitions hold for $\bar \sigma_i$ in terms of $\bar
\sigma_0$, $\bar \sigma_3$ and $\bar \sigma_8$. In Eq.~(\ref{semfa}) we
have also defined $\Sigma (p) = {\rm diag}( \Sigma_u(p), \Sigma_d(p),
\Sigma_s(p))$, with $\Sigma_i(p) = m_i + \bar \sigma_i\,r^2(p)$, whereas
the mean field values $\bar S_i$ are given by $\bar S_i = \tilde S_i (\bar
\sigma_j, 0)$. Note that in the isospin limit $\bar \sigma_u = \bar \sigma_d$,
thus we have $\Sigma_u(p) = \Sigma_d(p)$.

In order to deal with the mesonic degrees of freedom, we also
introduce a more convenient basis defined by
\begin{equation}
\phi_{ij} = \frac{1}{\sqrt2}\,\left(\lambda_a\phi_a\right)_{ij}\;,
\end{equation}
where $\phi=\sigma,\pi$, and the indices $i,j$ run from 1 to 3.
For the pseudoscalar fields one has in this way
\begin{equation}
\delta \pi_{ij} = \left( \begin{array}{ccc} \displaystyle
\frac{\pi^0}{\sqrt{2}} + \frac{\eta_8}{\sqrt{6}} +
\frac{\eta_0}{\sqrt3} & \pi^+ & K^+ \\ \pi^- & \displaystyle -
\frac{\pi^0}{\sqrt{2}} + \frac{\eta_8}{\sqrt{6}} +
\frac{\eta_0}{\sqrt3} & K^0 \\ K^- & \bar K^0 & \displaystyle -
\frac{2\,\eta_8}{\sqrt{6}} + \frac{\eta_0}{\sqrt3}
\end{array} \right)_{ij} .
\end{equation}
The second piece of the effective action in Eq.~(\ref{act})
---quadratic in the meson fluctuations--- can be written now as
\begin{equation}
S_E^{quad} = \frac{1}{2} \int \frac{d^4p}{(2\pi)^4} \left[
G^+_{ij,kl}(p) \ \delta\sigma_{ij}(p) \ \delta\sigma_{kl}(-p) +
G^-_{ij,kl}(p) \ \delta\pi_{ij}(p)\;\delta\pi_{kl}(-p) \right]\;,
\end{equation}
where we have defined
\begin{equation}
G^{\pm}_{ij,kl}(p) = C^{\pm}_{ij}(p) \ \delta_{il}\,\delta_{jk} +
\left((r^\pm)^{-1}\right)_{ij,kl}\;,
\end{equation}
with
\bea C^\pm_{ij}(p^2) & = &
 -\, 8 \,N_c \int \frac{d^4 q}{(2 \pi)^4}\ \frac{  r^2(q^+) r^2(q^-)
 \left[ (q^+ \cdot q^-) \mp \Sigma_i(q^+) \Sigma_j(q^-)\right]}{\left[
 (q^+)^2 + \Sigma_i^2(q^+) \right] \left[ (q^-)^2 +
 \Sigma_j^2(q^-)\right]}\;,\qquad q^\pm = q \pm p/2\;,
 \label{ciju} \\
r^\pm_{ij,kl} & = &  G  \ \delta_{il}\,\delta_{jk} \pm
\frac{H}{2}\ \epsilon_{ikn}\,\epsilon_{jln} \ \bar S_n \;. \eea

\subsection{Mean field approximation and chiral condensates}

The mean field values $\bar \sigma_u$ and $\bar\sigma_s$ can be
found by minimizing the action $S_E^{MFA}$. Taking into account
Eqs.~(\ref{spa}), a straightforward exercise leads to the
following set of coupled ``gap equations'':
\begin{equation}
\left\{\begin{array}{rcl} \displaystyle \bar \sigma_u + G\,\bar
S_u + \frac{H}{2} \, \bar S_u \bar S_s &=& 0 \\ \displaystyle \bar
\sigma_s + G\,\bar S_s + \frac{H}{2} \, \bar S_u^2 &=& 0\ \ ,
\end{array}\right.
\label{gapeq}
\end{equation}
where
\begin{equation}
\bar S_i = -\, 8\, N_c \int \frac{d^4 p}{(2 \pi)^4}\ \frac{
\Sigma_i(p) \ r^2(p) }{p^2 + \Sigma_i^2(p)}\;.
\end{equation}

The chiral condensates are given by the vacuum expectation values
$\langle\,\bar u u\,\rangle = \langle\,\bar d  d\,\rangle$ and
$\langle\,\bar s s\,\rangle$. They can be easily obtained by
performing the variation of $Z^{MFA} = \exp[ -S_E^{MFA} ]$ with
respect to the corresponding current quark masses. For $q=u,s$ one
obtains
\begin{equation}
\langle\,\bar q q\,\rangle = -\, 4\, N_c \int \frac{d^4 p}{(2
\pi)^4}\ \frac{\Sigma_q(p)}{p^2 + \Sigma_q^2(p)}\;.
\label{chiralcond}
\end{equation}

\subsection{Meson masses and quark-meson coupling constants}

{}From the quadratic effective action $S_E^{quad}$ it is possible
to derive the scalar and pseudoscalar meson masses as well as the
quark-meson couplings. In what follows we will consider explicitly
only the case of pseudoscalar mesons. The corresponding
expressions for the scalar sector are completely equivalent, just
replacing the upper indices ``$-$'' by ``$+$''. In terms of
physical fields, the contribution of the pseudoscalar mesons to
$S_E^{quad}$ can be written as
\begin{eqnarray}
\left. S_E^{quad}\right|_P & = & \frac12 \int
\frac{d^4p}{(2\pi)^4} \, \bigg\{ G_\pi (p^2) \left[ \pi^0 (p) \
\pi^0 (-p) + 2\,\pi^+ (p) \ \pi^- (-p)\right] \nonumber \\ & &
\qquad \qquad + G_K (p^2) \left[ 2\, K^0 (p) \ \bar K^0 (-p) + 2\,
K^+ (p) \ K^- (-p) \right] \nonumber \\ & & \qquad \qquad +
G_\eta(p^2) \ \eta(p) \ \eta(-p)+ G_{\eta'}(p^2) \ \eta'(p) \
\eta'(-p) \bigg\}\;. \label{quad}
\end{eqnarray}
Here, the fields $\eta$ and $\eta'$ are related to the $U(3)_V$
states $\eta_0$ and $\eta_8$ according to
\begin{eqnarray}
\eta &=& \cos \theta_\eta \ \eta_8 - \sin \theta_\eta\ \eta_0 \\ \eta'
&=& \sin \theta_{\eta'} \ \eta_8 + \cos \theta_{\eta'}\ \eta_0 \ ,
\end{eqnarray}
where the mixing angles $\theta_{\eta,\eta'}$ are defined in such a way
that there is no $\eta-\eta'$ mixing at the level of the quadratic action.
The functions $G_P(p^2)$ introduced in Eq.~(\ref{quad}) are given by
\bea
G_\pi(p^2) & = & \left[ (G + \frac{H}{2} \bar S_s)^{-1} + C^-_{uu}(p^2)
\right] \\
G_K (p^2) & = & \left[ (G + \frac{H}{2} \bar S_u)^{-1} +
C^-_{us}(p^2) \right] \\
G_{\eta} (p^2) & = & \frac{G^-_{88}(p^2) + G^-_{00}(p^2)}{2} - \sqrt{
\left[ G^-_{08}(p^2) \right]^2 \! + \! \left[ \frac{G^-_{88}(p^2) -
G^-_{00}(p^2)}{2} \right]^2} \\
G_{\eta'} (p^2) & = & \frac{G^-_{88}(p^2) + G^-_{00}(p^2)}{2} + \sqrt{
\left[ G^-_{08}(p^2) \right]^2 \! + \! \left[ \frac{G^-_{88}(p^2) -
G^-_{00}(p^2)}{2} \right]^2}\;,
\eea
where
\bea G^-_{88} (p^2) & = & \frac13 \left[ \frac{6 G - 4 H \bar S_u - 2 H
\bar S_s} {2 G^2 - G H \bar S_s - H^2 \bar S_u^2} + C^-_{uu}(p^2) +
2\,C^-_{ss}(p^2) \right] \nonumber \\
G^-_{08} (p^2) & = & \frac{\sqrt{2}}{3} \left[ \frac{H (\bar S_s - \bar
S_u)}{2 G^2 - G H \bar S_s - H^2 \bar S_u^2}
+ C^-_{uu}(p^2) - C^-_{ss}(p^2) \right] \nonumber \\
G^-_{00} (p^2) & = & \frac13 \left[ \frac{6 G + 4 H \bar S_u - H
\bar S_s}{2 G^2 - G H \bar S_s - H^2 \bar S_u^2} +
2\,C^-_{uu}(p^2) + C^-_{ss}(p^2) \right]\,.
\eea

The meson masses are obtained by solving the equations
\begin{equation}
G_P (-m_P^2) = 0\;, \label{gp}
\end{equation}
with $P=\pi$, $K$, $\eta$ and $\eta'$, while the $\eta$ and
$\eta'$ mixing angles, which are in general different from each
other, are given by
\begin{eqnarray}
\mbox{tan}\ 2\,\theta_{\eta,\eta'} &=& \frac{2\,
G^-_{08}(p^2)}{G^-_{00}(p^2) - G^-_{88}(p^2)}
\Big|_{p^2=-m^2_{\eta,\eta'}}\;.
\label{thetap}
\end{eqnarray}

Now the meson fields have to be renormalized, so that the residues of the
corresponding propagators at the meson poles are set equal to one. This
means that one should define renormalized fields $\tilde \phi(p) =
Z^{-1/2}_\phi \ \phi(p)$ such that, close to the poles, the quadratic
effective lagrangian reads
\begin{equation}
\left. {\cal L}_E^{quad}\right|_\phi  = \frac{1}{2} \left( p^2 +
m^2_\phi \right) \ \tilde \phi(p) \ \tilde \phi(-p)\;.
\end{equation}
In this way, the wave function renormalization constants $Z_P$ are
given by
\begin{equation}
Z_P^{-1}  = \frac{d G_P(p^2) }{dp^2} \bigg|_{p^2=-m_P^2}\;,
\label{zp1}
\end{equation}
with $P=\pi$, $K$, $\eta$ and $\eta'$. Finally, the meson-quark
coupling constants $G_{Pq}$ are given by the original residues of
the meson propagators at the corresponding poles,
\begin{equation}
G_{Pq} = Z^{1/2}_P\; .
\label{gpi}
\end{equation}

\subsection{Weak decay constants of pseudoscalar mesons}

By definition, the pseudoscalar meson weak decay constants are
given by the matrix elements of the axial currents $A^a_\mu(x)$
between the vacuum and the renormalized one-meson states at the
meson pole:
\begin{equation}
\langle\, 0 | A^a_\mu(0) | \tilde \phi_b (p) \,\rangle = i
\;f_{ab}\; p_\mu\;.
\end{equation}
For $a,b=1\dots 7$, the constants $f_{ab}$ can be written as
$\delta_{ab}\,f_\phi$, with $\phi = \pi$ for $a=1,2,3$ and
$\phi=K$ for $a=4$ to 7. In contrast, as occurs with the mass
matrix, the decay constants become mixed in the $a=0,8$ sector.

In order to obtain the expression for the axial current, one has
to gauge the effective action $S_E$ by introducing a set of axial
gauge fields ${\cal A}^a_\mu$. For a local theory, this gauging
procedure can be done simply by performing the replacement
\begin{equation}
\partial_\mu \rightarrow \partial _\mu + \frac{i}{2} \; \gamma_5
\; \lambda_a \; {\cal A}^a_\mu(x)\;.
\end{equation}
However, since in the present case we are dealing with
nonlocal fields, an extra replacement has to be performed in the regulator
$r(x-y)$~\cite{BB95,PB98,BKB91}. One has
\begin{equation}
r(x-y) \rightarrow \mbox{P} \exp \left[\frac{i}{2} \int_x^y ds_\mu
\ \gamma_5 \ \lambda_a \ {\cal A}^a_\mu(s) \right] r(x-y)\;,
\label{gaugreg}
\end{equation}
where $s$ represents an arbitrary path that connects $x$ with $y$.
In the present work we will use the so-called ``straight line
path'', which means
\begin{equation}
s_\mu = x_\mu + \alpha \left( y_\mu - x_\mu\right)\;,
\end{equation}
with $0 \le \alpha \le 1$. Once the gauged effective action is
obtained, it is straightforward to get the axial current as the
derivative of such action with respect to ${\cal A}^a_\mu(x)$
evaluated at ${\cal A}^a_\mu(x)=0$. Then, performing the
derivative of the resulting expressions with respect to the
renormalized meson fields, one can finally identify the
corresponding meson weak decay constants. After a rather lengthy
calculation, we find that the pion and kaon decay constants are
given by
\bea f_\pi &=& 4\, f_{uu}(-m_\pi^2)\  Z_\pi^{1/2}\;, \label{fpi}\\
f_K   &=& 2 \left[ f_{us}(-m_K^2) + f_{su}(-m_K^2) \right]
Z_K^{1/2}\;,
\eea
where
\begin{eqnarray}
f_{ij}(p^2) & = & 2\, N_c \int \frac{d^4 q}{(2 \pi)^4} \left. \
\frac{(p \cdot q_\alpha^+)}{p^2} \ \frac{ r(q_\alpha^+)
r(q_\alpha^-) \Sigma_j(q_\alpha^-)}{\left[(q_\alpha^+)^2 +
\Sigma_i^2(q_\alpha^+) \right]\left[ (q_\alpha^-)^2 +
\Sigma_j^2(q_\alpha^-) \right]}\;\right|_{\alpha=1/2} \nonumber \\
& & + \; 4\, N_c \int \frac{d^4 q}{(2 \pi)^4}\  \frac{(p \cdot
q)}{p^2} \ \frac{dr(q)}{d q^2}\, \int_0^1 d\alpha \
 \frac{1}{(q^+_\alpha)^2 + \Sigma_i^2(q^+_\alpha)} \nonumber \\
& &
 \times \ \bigg\{  r(q^-_\alpha)\; [\Sigma_i(q^+_\alpha)-m_i]\
 \frac{\left[ (q^+_\alpha \cdot q^-_\alpha) + \Sigma_i(q^+_\alpha)
 \Sigma_j(q^-_\alpha)\right]}{(q^-_\alpha)^2 + \Sigma_j^2(q^-_\alpha)}
+ \;r(q^+_\alpha)\; \Sigma_i(q^+_\alpha) \bigg\}\;, \label{fij}
\label{expfpi}
\end{eqnarray}
with \bea q^+_\alpha &=& q + (1-\alpha) \ p\;, \nonumber \\
q^-_\alpha &=& q - \alpha \ p \ . \eea

In the case of the $\eta-\eta'$ system, two decays constants can
be defined for each component $a=0,8$ of the axial
current~\cite{PDG02}. They can be written in terms of the $f_{ab}$
decay constants and the previously defined mixing angles
$\theta_{\eta,\eta'}$ as
\begin{eqnarray}
f^a_{\eta} &=& \left[ f_{a8}(-m_{\eta}^2) \cos \theta_{\eta} -
f_{a0}(-m_{\eta}^2) \sin \theta_{\eta}\right] \ Z_\eta^{1/2}
\label{faeta} \\
f^a_{\eta'} &=& \left[ f_{a8}(-m_{\eta'}^2) \sin \theta_{\eta'} +
f_{a0}(-m_{\eta'}^2) \cos \theta_{\eta'} \right] \
Z_{\eta'}^{1/2}\;.
\label{faetap}
\end{eqnarray}
Within our model, the decay constants $f_{ab}$ for $a,b=0,8$ are related
to the $f_{ij}$ defined in Eq.~(\ref{fij}) by
\bea f_{88}(p^2)  &=&
\frac{4}{3} \left[ 2 f_{ss}(p^2) + f_{uu}(p^2) \right] \\
f_{00}(p^2)  &=& \frac{4}{3} \left[ 2 f_{uu}(p^2) + f_{ss}(p^2) \right] \\
f_{08}(p^2) = f_{80}(p^2) &=& \frac{4\sqrt{2}}{3} \left[ f_{uu}(p^2) -
f_{ss}(p^2) \right]\;.
\label{f08}
\eea
It is clear that both the nondiagonal decay constants $f_{08}$, $f_{80}$
as well as the mixing angles $\theta_\eta$ and $\theta_{\eta'}$ vanish in
the $SU(3)$ symmetry limit.

\subsection{Anomalous $P\to\gamma\gamma$ decays}

To go further with the analysis of light pseudoscalar meson decays, let us
evaluate the anomalous decays of $\pi^0$, $\eta$ and $\eta'$ into two
photons. In general, the corresponding amplitudes can be written as
\begin{equation}
A(P\to\gamma\gamma) \ =
\ e^2\; g_{P\gamma\gamma}\ \epsilon_{\mu\nu\alpha\beta}
\ \varepsilon_1^{\ast\mu}\;\varepsilon_2^{\ast\nu} \;k_1^\alpha\;k_2^\beta\ ,
\end{equation}
where $P=\pi^0,\eta,\eta'$, and $k_i$, $\varepsilon_i$ stand for the
momenta and polarizations of the outgoing photons respectively.

In the nonlocal model under consideration the coefficients
$g_{P\gamma\gamma}$ are given by quark loop integrals. Besides the usual
``triangle'' diagram, given by a closed quark loop with one meson and two
photon vertices, in the present nonlocal scheme one has a second
diagram~\cite{PB98} in which one of the quark-photon vertices arises from
the gauge contribution to the regulator, see Eq.~(\ref{gaugreg}). The sums
of both diagrams for $\pi^0$, $\eta$ and $\eta'$ decays yield
\begin{eqnarray}
g_{\pi\gamma\gamma} & = & I_u(m_\pi^2)\ Z_\pi^{1/2} \nonumber \\
g_{\eta\gamma\gamma} & = & \frac{1}{3\sqrt3}\; \bigg[ \left[\,5\,
I_u(m_\eta^2) - 2\, I_s(m_\eta^2)\right]\cos\theta_\eta - \sqrt2\,
\left[\,5\, I_u(m_\eta^2) + I_s(m_\eta^2)\right]\sin\theta_\eta \bigg]
\ Z_\eta^{1/2} \nonumber \\
g_{\eta'\gamma\gamma} & = & \frac{1}{3\sqrt3}\; \bigg[ \left[\,5\,
I_u(m_{\eta'}^2) - 2\, I_s(m_{\eta'}^2)\right]\sin\theta_{\eta'} +
\sqrt2\, \left[\,5\, I_u(m_{\eta'}^2) +
I_s(m_{\eta'}^2)\right]\cos\theta_{\eta'} \bigg] \ Z_{\eta'}^{1/2}\ ,
\end{eqnarray}
where the loop integrals $I_f(m_P^2)$ are given by
\begin{eqnarray}
I_f(m_P^2) & = & \frac{8}{3}\; N_c \; \int\ \frac{d^4q}{(2\pi)^4}\ \
\frac{r(q-k_2)\;r(q+k_1)}
{[q^2+\Sigma_f^2(q)] \; [(q+k_1)^2+\Sigma_f^2(q+k_1)]
\; [(q-k_2)^2+\Sigma_f^2(q-k_2)]}\nonumber \\
& & \times \left\{ \Sigma_f(q)\ +\ \frac{q^2}{2}\;
\left[\frac{[\Sigma_f(q-k_2) - \Sigma_f(q)]}{(k_2\cdot q)} \ - \
\frac{[\Sigma_f(q+k_1) - \Sigma_f(q)]}{(k_1\cdot q)} \right] \right\}
\label{if}
\end{eqnarray}
(notice that for on-shell photons these integrals are only functions of
the scalar product $(k_1\cdot k_2)$, which in Euclidean space is equal to
$-m_P^2/2$). In terms of the parameters $g_{P\gamma\gamma}$, the
corresponding decay widths are simply given by
\begin{equation}
\Gamma(P\to \gamma\gamma) = \frac{\pi}{4}\; m_P^3\;\alpha^2\;
g_{P\gamma\gamma}^2\ ,
\end{equation}
where $\alpha$ is the fine structure constant.

\subsection{Low energy theorems}

Chiral models are expected to satisfy some basic low energy theorems. In
this subsection we consider some important relations such as the
Goldberger-Treiman (GT) and Gell-Mann-Oakes-Renner (GOR), showing
explicitly that they are indeed verified by the present model.

We start by the GT relation. Taking the chiral limit $m_\pi^2 \rightarrow
0$ in $f_{uu}(-m_\pi^2)$ and $dG_\pi/dp^2|_{p^2=-m_\pi^2}$ appearing in
Eqs.(\ref{fpi}) and (\ref{zp1}) we get
\begin{equation}
\lim_{m_\pi^2 \rightarrow 0} \ {f_{uu}(-m_\pi^2)} = \frac{1}{4} \
\bar\sigma_{u,0}\ Z_{\pi,0}^{-1}\ ,
\end{equation}
where here, as in the rest of this subsection, the subindex 0 indicates
that the corresponding quantity is evaluated in the chiral limit (notice
that $\bar\sigma_{u,0}=\Sigma_{u,0}(0)$). Replacing this expression in
Eq.~(\ref{fpi}) and taking into account Eq.~(\ref{gpi}) we get
\begin{equation}
f_{\pi,0} \  G_{\pi q,0} = \bar\sigma_{u,0}\ ,
\label{gtrelation}
\end{equation}
which is equivalent to the GT relation at the quark level in our model.

Let us turn to the GOR relation. Expanding $C^-_{uu}(p^2)$ in
Eq.~(\ref{ciju}) to leading order in $m_u$ and $p^2$ we get
\begin{equation}
C^-_{uu}(p^2)\; \simeq\; \frac{ \bar S_{u,0} }{ \bar \sigma_{u,0} }\; -\; m_u \;
\frac{  { \langle\; \bar u  u + \bar d d \;\rangle_0 }}
{\bar \sigma^2_{u,0}}\; +\; p^2 \; Z_{\pi,0}^{-1}\ .
\label{cuuexp}
\end{equation}
To obtain this result we have used the gap equations (\ref{gapeq}) and
the expression for the chiral condensate given in Eq.~(\ref{chiralcond}).
Now using Eq.~(\ref{cuuexp}) together with the equation for the pion mass,
\begin{equation}
G_\pi(- m_\pi^2) = 0\ ,
\label{pioneq}
\end{equation}
and taking into account the GT relation (\ref{gtrelation}), one gets
\begin{equation}
m_u \ \langle\; \bar u  u + \bar d  d \;\rangle_0 \ = - \ f_{\pi,0}^2 \
m_\pi^2\ ,
\label{GOR}
\end{equation}
which is the form taken by the well-known GOR relation in the isospin limit.

Next we discuss the validity of the Feynman-Hellman theorem for the case
of the so-called pion sigma term. This theorem implies the
relation~\cite{GAS81}
\begin{equation}
\frac{ d m_\pi^2 }{ d m_u } \ = \ \langle\; \pi\,|\;\bar u u + \bar d d
\;|\,\pi\; \rangle \ ,
\label{FH}
\end{equation}
where covariant normalization, $\langle p' | p\rangle = 2 E_p \ (2 \pi)^3
\ \delta^3(\vec p' - \vec p)$, has been used for the pion field. An
expression for the left hand side of Eq.~(\ref{FH}) can be easily obtained
by deriving Eq.~(\ref{pioneq}) with respect to the u quark mass. In fact,
\begin{equation}
0 = \frac{ d G_\pi (- m_\pi^2 ) }{d m_u} =
\left. \frac{ \partial G_\pi (p^2) }{\partial m_u} \right|_{p^2=-m_\pi^2} +
\left. \frac{ \partial G_\pi (p^2) }{\partial p^2} \right|_{p^2=-m_\pi^2} \
\left(- \frac{ d m_\pi^2 }{d m_u} \right)\ ,
\end{equation}
thus
\begin{equation}
\frac{ d m_\pi^2 }{d m_u} = \left. \frac{\displaystyle
\ \ \ \frac{\partial G_\pi (p^2)}{\partial m_u}\ \ \ }
{\displaystyle \frac{\partial G_\pi (p^2)}{\partial p^2}}
\ \right|_{p^2=-m_\pi^2}\ .
\label{pi2}
\end{equation}
On the other hand, within the path integral formalism, one has
\begin{equation}
\langle \; \pi^a \; | \; \bar u u + \bar d d \; | \; \pi^b \; \rangle =
\left. \frac{\delta^3 S^{j}_E[j]}{\delta j(0) \ \delta \pi^a(p) \ \delta
\pi^b(-p)} \right|_{{\footnotesize \begin{array}{l}
  \pi = j =0 \\
  p^2 = - m_\pi^2 \\
\end{array}}} \times \ Z_\pi\ ,
\end{equation}
where
\begin{equation}
S^{j}_E[j] = S_E + \int d^4x \ \left[ \ \bar u(x) u(x) + \bar d(x) d(x)\
\right] \ j(x)\ ,
\end{equation}
$S_E$ being the effective action of the model, given by Eq.~(\ref{se}).
{}From the explicit form of $S_E$ it is easy to see that
\begin{equation}
S^{j}_E[j] = S_E(m_u \rightarrow m_u + j(x))\ ,
\end{equation}
therefore, using the bosonized form of the effective action in
Eq.~(\ref{act}), with $S_E^{quad}$ given by Eq.~(\ref{quad}), we get
\begin{equation}
\langle\;\pi\; |\;\bar u u + \bar d d \; |\;\pi \rangle\; =\; Z_\pi
\left. \frac{\partial G^j_\pi[j]}{\partial j(0) }
\right|_{{\footnotesize \begin{array}{l}
  j =0 \\
  p^2 = - m_\pi^2 \\
\end{array}}}
=\;\frac{1}{ Z_\pi^{-1} } \ \left. \frac{\partial G_\pi(p^2)}{\partial
m_u} \right|_{p^2 = - m_\pi^2}\; = \;\left. \frac{ \displaystyle \frac{\
\partial G_\pi (p^2)\ }{\partial m_u} } {\displaystyle \frac{ \partial
G_\pi (p^2) }{\partial p^2}} \right|_{p^2=-m_\pi^2}\ .
\label{pibar}
\end{equation}
Comparing Eq.~({\ref{pibar}) with Eq.~({\ref{pi2}) we see that the FH
theorem, as it should, holds in the present model. Moreover, using the GOR
relation Eq.~(\ref{GOR}), we obtain up to leading order in $m_u$
\begin{equation}
\langle\;\pi\; |\;\bar u  u + \bar d  d\; |\;\pi\;\rangle_0 =
- \frac{ \langle\; \bar u  u + \bar d d\;\rangle_0}{f_{\pi,0}^2}
= \frac{ m_\pi^2 }{m_u}
\end{equation}

To conclude, let us analyze in the chiral limit the coupling
$g_{\pi\gamma\gamma}$. Expanding the integrand of $I_u(m_\pi^2)$ in powers
of $k_1,k_2$ and taking the limit $m_\pi^2\to 0$ one obtains~\cite{foot1}
\begin{equation}
g_{\pi\gamma\gamma,0}\ =\ \frac{Z_{\pi,0}^{1/2}}{4\pi^2\sigma_{u,0}}\ .
\label{relgpgg}
\end{equation}
Now, taking into account the GT relation, one finally has
\begin{equation}
g_{\pi\gamma\gamma,0}\ =\ \frac{1}{4\pi^2 f_{\pi,0}}\ ,
\label{FHrel}
\end{equation}
which is the expected result according to low energy theorems and Chiral
Perturbation Theory.

\section{Numerical results}

In this section we discuss the numerical results obtained within
the above described nonlocal model. Our results include the values
of meson masses, decay constants and mixing angles, as well as the
corresponding quark constituent masses, quark condensates and
quark-meson couplings. The numerical calculations have been
carried out for two different regulators often used in the
literature: the Gaussian regulator
\begin{equation}
r(p^2) = \exp{\left(-p^2/2\Lambda^2\right)}
\end{equation}
and a Lorentzian regulator
\begin{equation}
r(p^2) = \left(1 + p^2/\Lambda^2\right)^{-1}\ ,
\end{equation}
where $\Lambda$ is a free parameter of the model, playing the r\^ole of an
ultraviolet cut-off momentum scale. Let us recall that these regulators
are defined in Euclidean momentum space.

\subsection{Fits to physical observables}

The nonlocal model under consideration includes five free parameters.
These are the current quark masses $\bar m$ and $m_s$ ($\bar m=m_u=m_d$),
the coupling constants $G$ and $H$ and the cut-off scale $\Lambda$. In our
numerical calculations we have chosen to fix the value of $\bar m$,
whereas the remaining four parameters have been determined by requiring
that the model reproduces correctly the measured values of four physical
quantities. The observables we have used here are the pion and kaon
masses, the pion decay constant $f_\pi$, and a fourth quantity, chosen to
be alternatively the $\eta'$ mass or the squared $\eta'\to\gamma\gamma$
decay constant, $g_{\eta'\gamma\gamma}^2$. In the case of the Gaussian
(Lorentzian) regulator, we find that for $\bar m$ above a critical value
$m_{crit}\simeq 8.3$ MeV (3.9 MeV) the quark propagators have only complex
poles in Minkowski space. This can be understood as a sort of quark
confinement~\cite{BB95,PB98}. In contrast, for $m\lesssim m_{crit}$ one
finds that $u$ and $d$ quark Euclidean propagators do have at least two
doublets of purely imaginary poles (i.e.\ real poles in Minkowski space).

In Table I we quote our numerical results for several quantities of
interest. Besides the obtained values of meson masses and decay constants,
we include in this table the corresponding results for quark condensates,
constituent quark masses and quark-meson coupling constants. We have taken
into account both Gaussian and Lorentzian regulators, considering in each
case values for the light quark mass $\bar m$ above and below $m_{crit}$.
Sets GI, GIV, LI and LIV have been determined by fitting the free
parameters so as to reproduce the empirical value of $m_{\eta'}$, while
for sets GII, GIII, LII and LIII the $\eta'$ mass is obtained as a
theoretical prediction and the parameters have been determined by
adjusting $g_{\eta'\gamma\gamma}^2$ to its present central experimental
value. The last column of Table I shows the empirical values of masses and
decay constants to be compared with our predictions.

As stated in Sect.\ II.C, the meson masses are obtained by solving the
equations $G_P(-m_P^2)=0$ for $P=\pi$, $K$, $\eta$ and $\eta'$. Now, to
perform the corresponding numerical calculations, one has to deal with the
functions $C_{ij}(p^2)$ evaluated at Euclidean momentum $p^2=-m_P^2$.
These functions, defined by Eq.~(\ref{ciju}), include quark loop integrals
that need to be treated with special care when the meson mass exceeds a
given value $m_P > 2\,S_i$, where $S_i$ is the imaginary part (in
Euclidean space) of the first pole of the quark propagator. In practice
this may happen in the case of the $\eta'$ state, and physically it
corresponds to a situation in which the meson mass is beyond a
pseudothreshold of decay into a quark-antiquark state. A detailed
discussion on this subject is given in the Appendix. In particular, it is
seen that for $\bar m > m_{crit}$ (i.e.\ if the propagator has no purely
imaginary poles) the quark loop integrals can be regularized in such a way
that their imaginary parts vanish identically, and consequently the
$\eta'$ width corresponding to this unphysical decay is zero. In contrast,
for $\bar m < m_{crit}$ the width is in general nonzero, and the presence
of an imaginary part implies that the condition $G_{\eta'} (-m_{\eta'}^2)
= 0$ cannot be satisfied. In any case, it is still possible to define the
$\eta'$ mass by looking at the minimum of $|G_{\eta'} (-p_0^2)|$. The
situation is illustrated in Fig.\ 1, where we plot the absolute values of
the functions $G_P(-p_0^2)$ for Sets GI and GIV (upper and lower
panel, respectively). For Set GI ($\bar m > m_{crit}$) the $\bar uu$
pseudothreshold is reached at about 1 GeV, above the $\eta'$ mass, thus
the quark loop integrals are well defined and $G_{\eta'}(-m^2_{\eta'})=0$.
It can be seen that this is also the situation for the parameter Sets GII,
GIII, LII and LIII. On the other hand, for Set GIV ($\bar m < m_{crit}$,
lower panel in Fig.\ 1) the $\bar uu$ pseudothreshold is reached at $\sim
750$~MeV, well below the $\eta'$ mass. As stated, in this case one can
define $m_{\eta'}$ by looking at the minimum of the function
$|G_{\eta'}(-p_0^2)|$, which is represented by the dashed-dotted curve.
Though different from zero, the (unphysical) $\eta'$ width is relatively
small, and the minimum (which, as required, lies at $p_0=958$ MeV) is
close to the horizontal axis in the chosen GeV$^2$ scale. Therefore we do
not expect our results to be spoiled by confinement effects not included
in the model. In any case, we believe that for $\bar m < m_{crit}$ the
values of physical parameters related to $\eta'$ decay may not be
reliable, and they have not been included in Table I. A similar situation
occurs for Set LIV. Finally, in the case of Set LI it turns out that while
$\bar m > m_{crit}$ (no purely imaginary poles), the $\bar uu$
pseudothreshold occurs below the $\eta'$ mass. In this case, in order to
evaluate the loop integrals we have followed the regularization procedure
described in the Appendix, in which the imaginary parts of quark loop
integrals vanish, leading to a real $\eta'$ pole.

By examining Table I it can be seen that, for the chosen values of $\bar
m$, the results for the quark condensate $\langle \bar uu\rangle$, the
ratio $\langle \bar ss\rangle/\langle \bar uu\rangle$ and the constituent
quark masses $\Sigma_q(0)$ are similar to the values obtained within most
theoretical studies~\cite{reports,RW94}. The most
remarkable differences between the results corresponding to both
regulators are found in the values of the current quark masses and the
quark condensates. For the Gaussian regulator the parameters $\bar m$ and
$m_s$ are found to be about 8 MeV and 200 MeV respectively, while for the
Lorentzian regulator the corresponding values are approximately 4 MeV and
100 MeV. Note, however, that the obtained values of the ratio
$m_s/\bar m$ are very similar and somewhat above the present
phenomenological range $(m_s/m_u)^{emp} = 17-22$~\cite{PDG02}.
On the contrary, as expected from the GT relation and its generalizations,
quark condensates are found to be higher for the
Lorentzian regulator parameter sets. In any case, the results for both
regulators are in reasonable agreement with standard phenomenological
values~\cite{pheno} and the most recent lattice QCD
estimates~\cite{latQCD}.

The predicted values for the kaon decay constant are also
phenomenologically acceptable. Indeed, the prediction for the ratio
$f_K/f_\pi$ turns out to be significantly better than that obtained in the
standard NJL, where the kaon and pion decay constants are found to be
approximately equal to each other~\cite{reports} in contrast with
experimental evidence. It is also worth to notice that we obtain a very
good prediction for the $\pi^0\to\gamma\gamma$ decay rate. In this sense
the nonlocal model shows a further degree of consistency in comparison
with the standard local NJL model, where the quark momenta in the
anomalous diagrams should go beyond the cutoff limit in order to get a
good agreement with the experimental value~\cite{Rip97}.

Our results for $\eta$ and $\eta'$ masses and $\eta,\eta'\to\gamma\gamma$
anomalous decays require a more detailed discussion. In the Gaussian
regulator case, it is seen that Set GI, while fitting $m_{\eta'}$, leads
to a rather large value for $g_{\eta'\gamma\gamma}^2$. On the contrary,
for set GII, which fits the value of $g_{\eta'\gamma\gamma}^2$, the
$\eta'$ mass decreases up to a value of about 880 MeV (of course, one can
also choose intermediate sets between this two). In addition, it is found
that the fit to $g_{\eta'\gamma\gamma}^2$ has a second solution, namely
the parameter Set GIII, which leads to a $\eta'$ mass of about 1 GeV.
However, in this case $m_{\eta'}$ is found to be very close to the
pseudothreshold point, and as a consequence both the values of the
$\eta'$-quark coupling $G_{\eta'q}$ and the decay constant
$g_{\eta'\gamma\gamma}$ are quite sensible to small changes in the
parameters. On the other hand, for all four Gaussian regulator sets the
results for $m_\eta$ and $g_{\eta\gamma\gamma}$ do not change
significantly. The values for the $\eta$ mass are found to be in
relatively good agreement with experiment, while $g_{\eta\gamma\gamma}^2$
turns out to be somewhat large. In the case of the Lorentzian regulator,
the above described situation becomes strengthened: set LI leads to an
unacceptably large value for $g_{\eta'\gamma\gamma}$ (notice that, as
stated, here $m_{\eta'}$ is above the pseudothreshold), while set LII
leads to a too low $\eta'$ mass. Set LIII seems to reproduce reasonably
all measurable quantities, but as in the Gaussian regulator case, the
result for $g_{\eta'\gamma\gamma}$ is highly unstable. In general, as a
conclusion, one could say that the Gaussian regulator is preferred,
leading to a reasonable global fit of all quantities considered here.

It is worth to mention that the chosen value of $\bar m$ cannot be very
far from the values considered in Table I. For higher $\bar m$
one would obtain too low values for the quark condensates. On the other
hand, if $\bar m\gg m_{crit}$ the $\eta'$ mass turns out to be very
far above the $\bar uu$ pseudothreshold implying the existence of
a large unphysical $\eta'$ width.

\subsection{$\eta$ and $\eta'$ mixing angles and decay constants}

The problem of defining and (indirectly) fitting mixing angles and decay
constants for the $\eta-\eta'$ system has been revisited several times in
the literature (see Ref.~\cite{F00}, and references therein). On general
grounds one has to deal with two different state mixing angles $\theta_P$
and four decay constants $f_P^a$, where $P=\eta,\eta'$ and $a=0,8$.
Standard analyses, however, used to parameterize the mixing between both
meson states and decay constants using a single parameter (a mixing angle
usually called $\theta$), and introducing two decay constants, $f_8$ and
$f_0$, related to $g_{\eta\gamma\gamma}$ and $g_{\eta'\gamma\gamma}$
through low energy equations analogous to Eq.~(\ref{FHrel}). In the last
few years, the analysis has been improved (mainly in the framework of
next-to-leading order Chiral Perturbation Theory), and several authors
have considered the possible non-orthogonality of $(f_\eta^8,f_{\eta'}^8)$
and $(f_\eta^0,f_{\eta'}^0)$~\cite{L97,FK02}, as well as that of the
states $\eta$ and $\eta'$~\cite{EF99}. For the sake of comparison, we
follow here Ref.~\cite{L97} and express the four decays constants $f_P^a$
in terms of two decay constants $f_a$ and two mixing angles $\theta_a$,
where $a=0,8$. Namely,
\begin{equation}
\left(
\begin{array}{cc}
f_\eta^8 & f_\eta^0 \\ f_{\eta'}^8 & f_{\eta'}^0
\end{array}
\right) =
\left(
\begin{array}{cc}
f_8\,\cos\theta_8 & -f_0\,\sin\theta_0 \\
f_8\,\sin\theta_8 & \ \ f_0\,\cos\theta_0
\end{array}
\right)\ .
\label{fleut}
\end{equation}
In our model, the decay constants $f_P^a$ can be calculated from
Eqs.~(\ref{faeta}-\ref{f08}). As shown below, it turns out that the angles
$\theta_8$ and $\theta_0$ are in general different. This also happens with
the mixing angles $\theta_\eta$ and $\theta_{\eta'}$, whose expressions
are given in Eq.~(\ref{thetap}). Notice that these are consequences of the
(rather strong) $p^2$ dependence of the functions $C_{ij}(p^2)$ and
$f_{ij}(p^2)$ defined in Eqs.~(\ref{ciju}) and (\ref{expfpi}),
respectively.

Our numerical results for the the parameters $f_a$, $\theta_a$ introduced
in Eq.~(\ref{fleut}) and the mixing angles $\theta_{\eta,\eta'}$ are
collected in Table II. In the last column of the Table we quote the ranges
in which the parameters $f_a$, $\theta_a$ fall within most popular
theoretical approaches. We have taken these values from the analysis in
Ref.~\cite{F00}, in which the results from different parameterizations
have been translated to the four-parameter decay constant scheme given by
Eq.~(\ref{fleut}). By looking at Table II it is seen that the results
corresponding to Gaussian regulator sets lie within the range of values
quoted in the literature, while in the case of the Lorentzian regulator
the most remarkable difference corresponds to Set LI, which leads to a
large value of $f_0$ (this can be related with the unacceptably large
value of $g_{\eta'\gamma\gamma}$ discussed above). The mixing angles
$\theta_a$ can be compared to those obtained in a recent Bethe-Salpeter
approach calculation~\cite{NNTYO00}, which leads to a somewhat larger
(absolute) value of $\theta_0$. As stated, for Sets GIV and LIV the values
for the $\eta'$ decay constants have not been computed in view of the
unphysical finite width acquired by the $\eta'$ meson.

Finally, notice that for most parameter sets (both Gaussian and
Lorentzian) the mixing angle $\theta_\eta$ gets a small positive value,
whereas the angle $\theta_{\eta'}$ lies around $-45^\circ$. That is to
say, we find our mixing scheme to be very far from the approximation of a
single mixing angle $\theta$. For this reason we believe there is no reason
to expect $\theta_\eta$ to be within the usually quoted range
$\theta = - (10^\circ- 20^\circ)$.

\section{Conclusions}

In this work we have studied the properties of light pseudoscalar mesons
in a three flavor chiral quark model with nonlocal separable interactions,
in which the $U(1)_A$ breaking is incorporated through a nonlocal
dimension nine operator of the type suggested by 't Hooft. We consider the
situation in which the Minkowski quark propagator has poles at real
energies as well as the case where only complex poles appear, which has
been proposed in the literature as a realization of confinement. We
concentrate on the evaluation of the masses and decay constants of the
pseudoscalar mesons for two different nonlocal regulators, namely Gaussian
and Lorentzian.

As general conclusions, it is found that in this model the prediction for
the ratio $f_K/f_\pi$ turns out to be significantly better than that
obtained in the standard NJL, where the kaon and pion decay constants are
found to be approximately equal to each other in contrast with the
experimental evidence. In addition, the model overcomes the standard local
NJL problem of treating the anomalous quark loop integrals in a consistent
way. With respect to the $\eta-\eta'$ system, according to our analysis
the Gaussian regulator seems to be more adequate than the Lorentzian one
to reproduce the observed phenomenology. In general, our results are in
reasonable agreement with experimentally measured values, and the global
fits can be still improved by considering regulators of more sophisticated
shapes. Alternatively, one might consider adding degrees of freedom not
explicitly included in the present calculations such as explicit vector
and axial-vector interactions, two-gluon components for the $\eta$ and
$\eta'$ mesons, etc. Finally, it is worth to remark that our fits lead to
an $\eta-\eta'$ system in which the $U(3)$ states $\eta_8$ and $\eta_0$
are mixed by two angles $\theta_\eta$ and $\theta_{\eta'}$ that appear to
be significantly different from each other.

\section{Acknowledgements}
We thank M. Birse and R. Plant for useful correspondence. NNS also would
like to acknowledge discussions with P. Faccioli and G. Ripka. This work
was partially supported by ANPCYT (Argentina) grant
PICT00-03-0858003(NNS). The authors also acknowledge support from
Fundaci\'on Antorchas (Argentina).

\pagebreak

\section*{Appendix: Evaluation of quark loop integrals}

\newcounter{erasmo}
\renewcommand{\thesection}{\Alph{erasmo}}
\renewcommand{\theequation}{\Alph{erasmo}.\arabic{equation}}
\setcounter{erasmo}{1} \setcounter{equation}{0} 

In this Appendix we describe some details concerning the evaluation of the
quark loop integrals $C_{ij}(p^2)$ defined in Eq.~(\ref{ciju}). Notice
that in the nonlocal chiral model analyzed in this work all four-momenta
are defined in Euclidean space. However, in order to determine the meson
masses, the external momenta $p$ in the loop integrals have to be extended
to the space-like region. Hence, without loss of generality, we choose
$p=(\vec 0, i p_0)$, and use three-momentum rotational invariance to write
the quark loop integrals as
\begin{equation}
C_{ij}(-p_0^2) = \int dq_3 \ dq_4 \ q_3^2 \frac{ F_{ij}(q_3,q_4,p_0) }
{\left[ (q^+)^2 + \Sigma_i^2(q^+) \right] \left[ (q^-)^2 +
\Sigma_j^2(q^-) \right]}\ ,
\label{cij}
\end{equation}
where $q_3 = |\vec q|$. The explicit form of $F_{ij}(q_3,q_4,p_0)$, which
depends on the meson under consideration, can be easily obtained by
comparing Eqs.~(\ref{cij}) and (\ref{ciju}). In principle, the integration
in Eq.~(\ref{cij}) has to be performed over the half-plane $q_3
\,\epsilon\, [\,0,\infty)$, $q_4\,\epsilon\, (-\infty,\infty)$. For
sufficiently small values of $p_0$ the denominator does not vanish at any
point of this integration region. However, when $p_0$ increases it might
happen that some of the poles of the integrand pinch the integration
region, making the loop integral divergent. In such cases one has to find
a way to redefine the integral in order to obtain a finite result. In
practice, we need to extend the calculation of the loop integrals
$C_{ij}(-p_0^2)$ to relatively large values of $p_0$ only when trying to
determine the $\eta'$ mass. For this particle, one only has to deal with
integrals in which $i=j$, therefore we will restrict here to this case and
the indices $i,j$ will be dropped from now on.

Let us start by analyzing the zeros of the denominator in the integrand in
Eq.~(\ref{cij}). This denominator can be written as $D=D^+ D^-$, where
\begin{equation}
D^\pm = (q^\pm)^2 + \Sigma^2(q^\pm) \;.
\end{equation}
It is clear that the zeros of $D$ are closely related to the poles of the
quark propagator $S(q) = [-\rlap/q + \Sigma(q)]^{-1}$. In what follows we
will assume that the regulator is such that the propagator has a numerable
set of poles in the complex $q$ plane, and that there are no cuts. It is
not hard to see that these poles appear in multiplets that can be
characterized by two real numbers $(S_r^\nu,S_i^\nu)$, with $S_r^\nu \ge
0$, $S_i^\nu > 0$. The index $\nu\,\epsilon\,\mathbb{N}$ has been
introduced to label the multiplets, with the convention that $\nu$
increases for increasing $S_i^\nu$. It is convenient to distinguish
between two different situations: (a) there are some purely imaginary
poles, i.e.\ there are one or more $\nu$ for which $S_r^\nu =0$; (b) no
purely imaginary pole exists, i.e.\ $S_r^\nu > 0$ for all $\nu$. It can be
shown that purely imaginary poles show up as doublets located at Euclidean
momentum $(\sqrt{q^2})^\nu = \pm\, i\, S^\nu_i$, while complex
poles~\cite{foot2} appear as quartets located at $(\sqrt{q^2})^\nu =
S^\nu_r \pm i \ S^\nu_i$ and $(\sqrt{q^2})^\nu = - S^\nu_r \pm i \
S^\nu_i$. Clearly, the number and position of the poles depend on the
specific shape of the regulator. For the Gaussian interaction, three
different situations might occur. For values of $\bar \sigma$ below a
certain critical value $\bar \sigma_{c}$, two pairs of purely imaginary
simple poles and an infinite set of quartets of complex simple poles
appear. It is possible to check that in this case one of the purely
imaginary doublets is the multiplet which has the smallest imaginary part
($\nu=1$, according to our convention). At $\bar \sigma = \bar
\sigma_{c}$, the two pairs of purely imaginary simple poles turn into a
doublet of double poles with $S_r=0$, while for $\bar \sigma > \bar
\sigma_{c}$ only an infinite set of quartets of complex simple poles is
obtained. In the case of the Lorentzian interactions, there is also a
critical value above which purely imaginary poles at low momenta cease to
exist. However, for this family of regulators the total number of poles is
always finite.

As stated, for low enough external momentum $p_0$ the integrand in
Eq.~(\ref{cij}) does not diverge along the integration region. As $p_0$
increases, the first set of poles to be met is that with the lowest value
of $S_i$, namely $(\sqrt{q^2})^{\nu=1}$. In the calculation of the meson
properties mentioned in the main text, we deal with relatively low
external momenta, so that the effect of higher poles is never observed.
Thus, in order to simplify the discussion we will only consider in what
follows the first pole multiplet, dropping the upper index $\nu$. The
extension to the case in which other sets of poles become relevant will be
briefly commented at the end of this Appendix.

The denominator $D$ vanishes when $D^+=0$ and/or $D^-=0$, i.e.\ when
\begin{equation}
(q^+)^2 = q^2_3 + q^2_4 - \frac{p_0^2}{4} + i q_4 p_0 = S_r^2 - S_i^2
\pm 2 i S_r S_i
\end{equation}
and/or
\begin{equation}
(q^-)^2 = q^2_3 + q^2_4 - \frac{p_0^2}{4}
- i q_4 p_0 = S_r^2 - S_i^2 \pm 2 i S_r S_i\ .
\end{equation}
Solving these equations for $q_4$ we get in general eight different
solutions. Four of them are given by
\bea
q_4^{(3,1)} & = & -\frac{S_i S_r}{\gamma(q_3,S_i,S_r)}
+ i \left( \pm  \gamma(q_3,S_i,S_r) - \frac{p_0}{2} \right)
\label{p3s} \\
q_4^{(4,2)} & = & -\frac{S_i S_r}{\gamma(q_3,S_i,S_r)} + i
\left( \pm  \gamma(q_3,S_i,S_r) + \frac{p_0}{2} \right)\;,
\label{p4s}
\eea
where
\begin{equation}
\gamma(q_3,S_i,S_r) =  \sqrt{ \frac{ q_3^2 + (S_i^2 - S_r^2) +
\sqrt{ q_3^4 + 2 q_3^2 (S_i^2 - S_r^2) + (S_i^2 + S_r^2)^2}}{2}}\;,
\end{equation}
and the other four solutions are ${q_4^{(i)}}' = -\,{\rm Re}(q_4^{(i)}) +
i \,{\rm Im}(q_4^{(i)})$, with $i=1\dots 4$. In Eqs.\ (\ref{p3s}) and
(\ref{p4s}) $q_4^{(3,1)}$ correspond to the zeros of $D^+$ and
$q_4^{(4,2)}$ to those of $D^-$, while a similar correspondence holds for
${q_4^{(i)}}'$. For purely imaginary poles one has $S_r=0$, hence only
four independent solutions exist.

If $p_0$ is relatively small, the distribution of the poles in the complex
$q_4$ plane is that represented in Fig.\ \ref{figa1}. Fig.\ \ref{figa1}a
holds for situation (a), in which the poles in the first multiplet are
purely imaginary, while Fig.\ \ref{figa1}b corresponds to case (b), in
which these poles are complex. In both figures the dots indicate the zeros
of $D^+$ and the squares those of $D^-$. As we see, for small values of
$p_0$ half of the poles of $D^+$ (namely, $q_4^{(1)}$ for case (a) and
$q_4^{(1)}$ and ${q_4^{(1)}}'$ for case (b)) are placed below the real
axis, whereas the other half ($q_4^{(3)}$ for case (a) and $q_4^{(3)}$ and
${q_4^{(3)}}'$ for case (b)) lie above it. Something similar happens for
the poles of $D^-$. Now, as $p_0$ increases, the poles move in the
direction indicated by the arrows. For a certain value of $p_0$, the poles
$q_4^{(2)}$ and $q_4^{(3)}$ meet on the real $p_4$ axis (the same
obviously happens with ${q_4^{(2)}}'$ and ${q_4^{(3)}}'$ in case (b)),
thus pinching the integration region of the $(q_3,q_4)$ integral in
Eq.~(\ref{cij}). The location of this so-called ``pinch point'' in the
$(q_3,q_4)$ plane, which we denote by $(q^p_3,q^p_4)$, is given by the
solution of ${\rm Im}\,q_4^{(2)} = {\rm Im}\,q_4^{(3)} = 0$:
\begin{equation}
(q^p_3,q^p_4) = \left( \frac{ \sqrt{ \left( p_0^2 - 4 S_i^2 \right)
\left( p_0^2 + 4 S_r^2 \right)}} {2 p_0} , \pm \frac{2 S_i S_r}{p_0}
\right)\;.
\label{pinch}
\end{equation}
As we see, for $p_0 < 2 S_i$ there is no pinch point (actually, it occurs
for a complex value of $q_3$, outside the integration region in
Eq.~(\ref{cij})), while for $p_0 \ge 2 S_i$ one or two pinch points exist
depending on whether $S_r=0$ (case (a)) or $S_r\neq 0$ (case (b)). In this
way, for $p_0 > 2\,S_i$ the integral in Eq.~(\ref{cij}) turns out to be
ill-defined.

In order to find a proper regularization procedure, let us analyze a
simpler situation in which the problem might be solved in Minkowski space
through the usual ``$i\,\epsilon$'' prescription. We consider the loop
integral that appears in the usual Nambu-Jona-Lasinio model with
three-momentum cut-off,
\begin{equation}
I_2(-p_0^2) = 2\, i \lim_{\epsilon\rightarrow 0^+} \int^{\Lambda_3}
\frac{d^4 q_M}{(2 \pi)^4} \frac{1} {\left[(q_M^+)^2 - m^2 + i \epsilon
\right] \left[(q_M^-)^2 - m^2 + i \epsilon \right]}\;,
\label{i2}
\end{equation}
where we have added the subscript $M$ to stress the fact that here the
momenta are defined in Minkowski space. For sufficiently small values of
$p_0$, even in the limit $\epsilon\rightarrow 0^+$, the integral is
convergent and no regularization is needed. Thus one can simply perform
the Wick rotation $p_4=i\, p_0$ and take the limit $\epsilon\rightarrow
0^+$ even before performing the integration. One gets in this way
\begin{equation}
I_2(p_4^2) = - 2 \int^{\Lambda_3} \frac{d^4 q_E}{(2 \pi)^4} \frac{1}
{\left[(q_E^+)^2 + m^2 \right] \left[(q_E^-)^2 + m^2 \right]}\;,
\label{i2E}
\end{equation}
which is an integral of the type given in Eq.~(\ref{cij}). Note that the
poles of the propagators are such that this situation belongs to case (a),
with $S_i=m$. For $p_0 > 2\, m$ the straightforward transformation from
Minkowski to Euclidean space mentioned above cannot be done, since some
poles go through the integration contours. The question is whether the
result of the well-defined Miskowskian integral (\ref{i2}) can be still
recovered if one starts with the Euclidean integral (\ref{i2E}), which is
ill-defined for $p_0 > 2\, m$ due to the presence of a pinch point at
$(q_3^p,q_4^p) = (\sqrt{p_0^2/4 - m^2},0)$. It is not hard to prove that
the answer is yes, once the $q_4$ integration contours and the pole
positions are conveniently modified. The procedure requires to introduce
two small parameters, $\epsilon$ and $\delta$, and take the limit
$\delta\to 0^+$, $\epsilon\to 0^+$ at the end of the calculation. The
parameter $\epsilon$ is used to shift the poles of $D^+$ and $D^-$ (see
Fig.\ \ref{figa2}), whereas $\delta$ is used to split the $q_3$
integration interval in three subintervals: the first region corresponds
to $q_3 > q_3^p+\delta$, the second to $q_3^p -\delta< q_3 < q_3^p
+\delta$ and the third to $q_3 < q_3^p -\delta$. For each $q_3$ region we
define a different $q_4$ integration contour, as represented in Fig.\
\ref{figa2} (in the second region, Fig.\ 3b, also an arbitrary constant
$\kappa > 1$ is introduced). In fact, in the first and third regions the
limit $\epsilon \rightarrow 0^+$ can be taken even before performing the
integrations. These two $q_3$ regions give the full contribution to the
real part of the result. However, more care has to be taken with the
intermediate region, which is responsible for the full contribution to the
imaginary part. For example ---as it is well known from the
``$i\,\epsilon$'' Minkowskian formulation---, changing the sign of
$\epsilon$ does not affect the real part of the result but does change the
sign of the imaginary part.

The prescription just described can be now applied to regularize any loop
integral of the form given in Eq.~(\ref{cij}) in which $p_0 > 2 S_i$. Let
us consider first the case (a), for which the lowest set of poles has
$S_r=0$ and the pinch point is located at $(q_3^p,q_4^p) =
(\sqrt{p_0^2/4-S_i^2},0)$. Defining $C^{(a)}(-p_0^2)$ as an integral of
the form given by Eq.~(\ref{cij}) for which only one set of purely
imaginary poles contributes, we get
\bea
{\rm Re}\left[C^{(a)}(-p_0^2)\right] & = & \lim_{\delta\rightarrow 0^+}
\left\{ \; R^{(a)}(-p_0^2,\delta) +
\int_0^{q_3^p-\delta} dq_3 \int_{-\infty}^\infty dq_4
\frac{ q_3^2 \ F(q_3,q_4,p_0) } {\left[ (q^+)^2
+ \Sigma^2(q^+) \right] \left[ (q^-)^2 + \Sigma^2(q^-)\right]}
\right. \nonumber \\
& & \ \ + \left. \int_{q_3^p+\delta}^\infty \ dq_3 \int_{-\infty}^\infty
dq_4 \frac{ q_3^2 \ F(q_3,q_4,p_0) } {\left[ (q^+)^2 + \Sigma^2(q^+)
\right] \left[ (q^-)^2 + \Sigma^2(q^-)\right]} \; \right\} \label{real}
\\
{\rm Im}\left[C^{(a)}(-p_0^2)\right] & = & - \frac{\pi^2}{2 p_0}\;\
\frac{q_3^p \ F(q_3^p,0,p_0)}{
\left[ 1 + \left. \frac{ \partial \Sigma^2(q)}{\partial
q^2}\right|_{q^2=-S^2_i} \right]^2 } \ \ .
\label{imag}
\eea
Here $R^{(a)}(-p_0^2,\delta)$ is the so-called ``residue contribution'',
responsible for the cancellation of the divergence appearing in the
integrals in (\ref{real}) in the limit $\delta\to 0^+$. Its explicit
expression reads
\begin{equation}
R^{(a)}(-p_0^2,\delta) \  = \ 2\pi \int_0^{q_3^p-\delta}
\frac{dq_3}{\sqrt{q_3^2+S_i^2}} \ \ {\rm Re}\! \left[ \frac{ q_3^2\;
F(q_3,q_4,p_0)} {\left[ 1 + \frac{
\partial \Sigma^2(q^-)}{\partial (q^-)^2 } \right] \left[ (q^+)^2 +
\Sigma^2(q^+) \right] }
\right]_{q_4 = q_4^{(2)}= i(\frac{p_0}{2}-\sqrt{q_3^2+S_i^2})} \ .
\end{equation}
For case (b) we have to extend the previous analysis to the situation in
which the poles are complex even in the limit $\epsilon\to 0^+$. In this
case there is an ambiguity on how to extend the ``$i\,\epsilon$''
prescription already in Minkowski space. Here we will follow the
suggestion made in Ref.~\cite{CLOP69}, in which opposite signs of
$\epsilon$ are used for each pole and its hermitian conjugate (both
defined in Minkowski space). In our Euclidean framework, this corresponds
to choose different signs of $\epsilon$ for sets $p_4^{(i)}$ and
${p_4^{(i)}}'$. It is not hard to see that with this prescription the
contributions to the imaginary part of the quark loop integral coming from
both sets of poles cancel each other. In this way, defining
$C^{(b)}(-p_0^2)$ as an integral of the type given in Eq.~(\ref{cij}) for
which only one set of complex poles contributes, we get
\begin{equation}
{\rm Im}\left[C^{(b)}(-p_0^2)\right] = 0\ .
\end{equation}
For the real part one obtains an expression similar to Eq.~(\ref{real}),
just replacing $R^{(a)}(-p_0^2,\delta)$ by $R^{(b)}(-p_0^2,\delta)$, with
\begin{equation}
R^{(b)}(-p_0^2,\delta) \  = \  4 \pi \int_0^{q_3^p-\delta} \! \! dq_3 \ \
{\rm Re}\! \left[ \frac{ q_3^2\ F(q_3,q_4,p_0)} {\left[ 1 + \frac{
\partial \Sigma^2(q^-)}{\partial (q^-)^2 } \right] \left[ (q^+)^2 +
\Sigma^2(q^+) \right] (i q_4 + \frac{p_0}{2}) } \right] _{q_4 = q_4^{(2)}}
\ .
\end{equation}

In principle, the extension of the present analysis to the situation in
which further sets of poles are relevant is rather straightforward.
However, some care has to be taken if one has more than one set of purely
imaginary poles, since in that case double poles might show up for
$p_0 > S_i^1 + S_i^2$.

One is faced with a similar problem in the calculation of the
$\eta'\to\gamma\gamma$ decay constant $g_{\eta'\gamma\gamma}$, where the
loop integral $I_u(m_{\eta'}^2)$ defined by Eq.~(\ref{if}) is divergent
for $m_{\eta'}>2\,S_i$. Though the situation is slightly more involved
(one finds two pinch points instead of one), our regularization
prescriptions can be trivially extended to include this case. In the same
way, the method could be extended to more complicated situations, as e.g.\
those found in Schwinger-Dyson type of calculations~\cite{RW94,BPT03}.
Finally, we note that although our prescription has some similarities with
that used in Ref.~\cite{PB98}, the regularization procedure is not exactly
the same. In the $(q_3,q_4)$ integral which appears in Eq.~(\ref{real})
the excluded region around the pinch point is a slide of infinite size in
the $q_4$ direction and size $2 \delta$ in the $q_3$ direction, as opposed
to the circular region of radius $\delta$ used in Ref.~\cite{PB98}. This
leads to some minor differences in the numerical values of the regulated
integrals.

\pagebreak

\begin{table}
\begin{tabular}{cccccccccccc} \hline
& & \multicolumn{4}{c}{Gaussian} & & \multicolumn{4}{c}{Lorentzian} &
\\
\hline
Set                &
        & GI & GII & GIII & GIV & & LI & LII & LIII & LIV & Empirical  \\
\hline
$\bar m$ &  \mbox{\esp}[ MeV ]\mbox{\esp}
         & \mbox{\esp}8.5\mbox{\esp} & \mbox{\esp}8.5\mbox{\esp} &
           \mbox{\esp}8.5\mbox{\esp} & \mbox{\esp}7.5\mbox{\esp} & \esp &
           \mbox{\esp}4.0\mbox{\esp} & \mbox{\esp}4.0\mbox{\esp} &
           \mbox{\esp}4.0\mbox{\esp} & \mbox{\esp}3.5\mbox{\esp} &
           \mbox{\esp\esp}(3.4\ -\ 7.4)\mbox{\esp\esp} \\
\hline
$m_s$               &  [ MeV ]
        & 223 & 223 & 223 & 199 & & 112 & 110 & 112 & 100 & (108\ -\ 209)    \\
$\Lambda$           &  [ MeV ]
        & 709 & 709 & 709 & 768 & & 1013 & 1013 & 1013 & 1110 &          \\
$G\,\Lambda^2$       &
        & 10.99 & 11.44 & 10.80 & 10.43 & & 14.68 & 16.76 & 14.96 & 14.05 & \\
$-H\,\Lambda^5$      &
        & 295.3 & 275.4 & 303.7 & 305.1 & & 743.4 & 573.8 & 720.4 & 821.0 & \\
\hline
-$<\bar u u>^{1/3}$ &  [ MeV ]
        & 211 & 211 & 211 & 220 & & 275 & 275 & 275 & 288 &  \\
-$<\bar s s>^{1/3}$ &  [ MeV ]
        & 186 & 187 & 185 & 204 & & 297 & 307 & 299 & 314 &  \\
$\Sigma_u(0)$       &  [ MeV ]
        & 313 & 313 & 313 & 295 & & 299 & 299 & 300 & 281 &  \\
$\Sigma_s(0)$       &  [ MeV ]
        & 650 & 662 & 645 & 607 & & 562 & 615 & 569 & 518 &  \\ \hline
$m_\pi$             &  [ MeV ]
& \ $139^*$ & \ $139^*$ & \ $139^*$ & \ $139^*$ & & \ $139^*$
& \ $139^*$& \ $139^*$ & \ $139^*$ &  139  \\
$m_K$               &  [ MeV ]
& \ $495^*$ & \ $495^*$ & \ $495^*$ & \ $495^*$ & & \ $495^*$
& \ $495^*$ & \ $495^*$ & \ $495^*$ &  495  \\
$m_\eta$            &  [ MeV ]
        & 517 & 505 & 521 & 522 & & 543 & 513 & 540 & 545 &  547   \\
$m_{\eta'}$         &  [ MeV ]
& \ $958^*$ & 879 & 1007 & \ $958^*$ & & \ $958^*$ & 778 & 908 & \ $958^*$ & 958 \\
\hline
$G_{\pi q}$         &
        & 3.28 & 3.28 & 3.28 & 3.09 & & 3.13 & 3.13 & 3.13 & 2.94 &     \\
$G_{K q}$           &
        & 3.47 & 3.52 & 3.45 & 3.21 & & 3.05 & 3.24 & 3.08 & 2.80 &     \\
$G_{\eta q}$        &
        & 3.07 & 3.03 & 3.08 & 2.83 & & 2.74 & 2.69 & 2.74 & 2.49 &     \\
$G_{\eta' q}$       &
        & 1.62 & 2.01 & 1.21 &      & & 1.36 & 2.21 & 1.13 &      &     \\
\hline
$f_\pi$             &  [ MeV ]
& \ $93.3^*$ & \ $93.3^*$ & \ $93.3^*$ & \ $93.3^*$ & & \ $93.3^*$
& \ $93.3^*$ & \ $93.3^*$ & \ $93.3^*$ & 93.3  \\
$f_K/f_\pi$         &
        & 1.29 & 1.29 & 1.29 & 1.29 & & 1.25 & 1.28 & 1.26 & 1.25 & 1.22  \\
$g_{\pi\gamma\gamma}^2$   &  [ GeV$^{-2}$ ]
& 0.073 & 0.073 & 0.073 & 0.074 & & 0.074 & 0.074 & 0.074 & 0.074 & $0.075\pm 0.005$ \\
$g_{\eta\gamma\gamma}^2$  &  [ GeV$^{-2}$ ]
& 0.095 & 0.106 & 0.091 & 0.094 & & 0.072 & 0.108 & 0.075 & 0.075 & $0.067\pm 0.006$ \\
$g_{\eta'\gamma\gamma}^2$ &  [ GeV$^{-2}$ ]
& 0.141 & \ $0.116^*$ & \ $0.116^*$ &     &
& 0.278 & \ $0.116^*$ & \ $0.116^*$ &     & $0.116\pm 0.005$ \\
\hline
\end{tabular}
\caption{Numerical results for quark effective masses and condensates, and
pseudoscalar meson masses and decay parameters. The light quark mass $\bar
m$ has been taken as input, and the parameters $m_s$, $\Lambda$, $G$ and
$H$ have been chosen so as to reproduce the empirical values of pion and
kaon masses, the pion decay constant $f_\pi$ and, alternatively, the
$\eta'$ mass or the measured value of $g_{\eta'\gamma\gamma}^2$ (marked
with $^*$).
\vspace*{1cm}
\label{tab1}}
\end{table}

\begin{table}
\begin{tabular}{ccccccccccc} \hline
Set      & GI & GII & GIII & GIV & & LI & LII & LIII & LIV &
$\begin{array}{c} {\rm Theory\ \ \&} \\ {\rm Phenomenology} \end{array}$ \\
\hline
$f_8/f_\pi$
   & 1.33 & 1.34 & 1.32 & & & 1.32 & 1.36 & 1.31 &  &
           \mbox{\esp\esp\esp}(1.2\ -\ 1.4)\mbox{\esp\esp\esp} \\
$f_0/f_\pi$
   & 1.28 & 1.18 & 1.18 & & & 1.64 & 1.18 & 1.05 &  &
           \mbox{\esp\esp\esp}(1.0\ -\ 1.3)\mbox{\esp\esp\esp} \\
\hline
$\theta_8$
   & $-14.2^\circ$\ \ \ & $-19.7^\circ$\ \ \ & $-10.0^\circ$\ \ \ &  &
   & $-11.7^\circ$\ \ \ & $-25.4^\circ$\ \ \ & $-10.6^\circ$\ \ \ &
   & $-$($22^\circ$\ - $19^\circ$)\ \ \ \\
$\theta_0$
   & $-2.14^\circ$\ \ \ & $-5.25^\circ$\ \ \ & $-1.17^\circ$\ \ \ &  &
   & \ $1.66^\circ$ & $-6.52^\circ$\ \ \ & \ 1.60$^\circ$ &
   & $-$($10^\circ$\ -\ $0^\circ$)\ \ \ \\
\hline
$\theta_\eta$
   & 4.65 & 1.81 & 5.69 & 5.32 & & 7.69 & $-0.72$\ \ \ & 6.90 & 8.00 & \\
$\theta_{\eta'}$
   & $-50.0$\ \ \ & $-47.1$\ \ \ & $-52.4$\ \ \ & &
   & $-48.4$\ \ \ & $-44.6$\ \ \ & $-50.6$\ \ \ & & \\
\hline
\end{tabular}
\caption{Numerical results for $\eta$ and $\eta'$ weak decay parameters and
state mixing angles, according to the definitions given by Eqs.~(\ref{thetap})
and (\ref{fleut}).
\label{tab2}}
\end{table}

\pagebreak

\begin{figure}
\begin{center}
\centerline{\psfig{figure=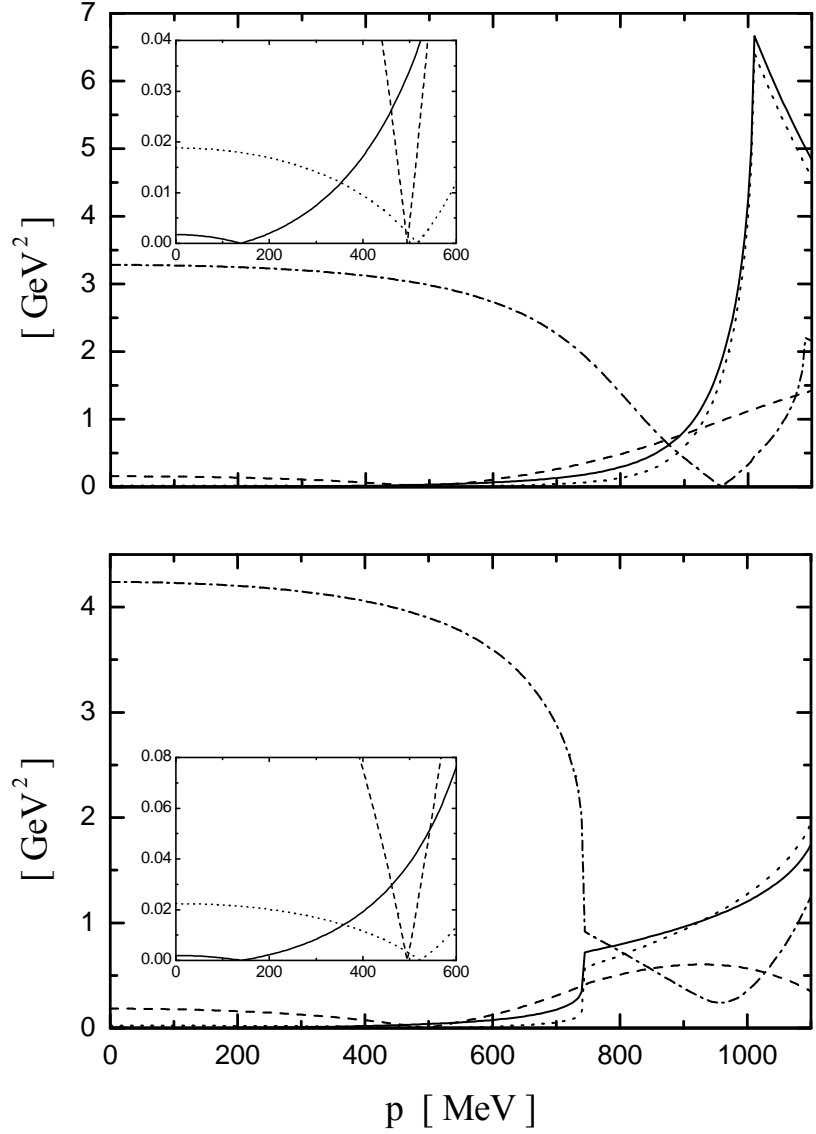,height=18cm}}
\caption{Absolute values of the inverse meson propagators as functions of
the momentum. The full line corresponds to $G_\pi$, the dashed line to
$G_K$, the dotted line to $G_\eta$ and the dashed-dotted line to
$G_{\eta'}$. The upper panel corresponds to Set GI and the lower one to
Set GIV. In both cases the insertions show a detail of the low momentum
region where the zeros corresponding to the ground state pion, kaon and
eta mesons occur. \label{fig1}}
\end{center}
\end{figure}

\pagebreak

\begin{figure}
\begin{center}
\centerline{\psfig{figure=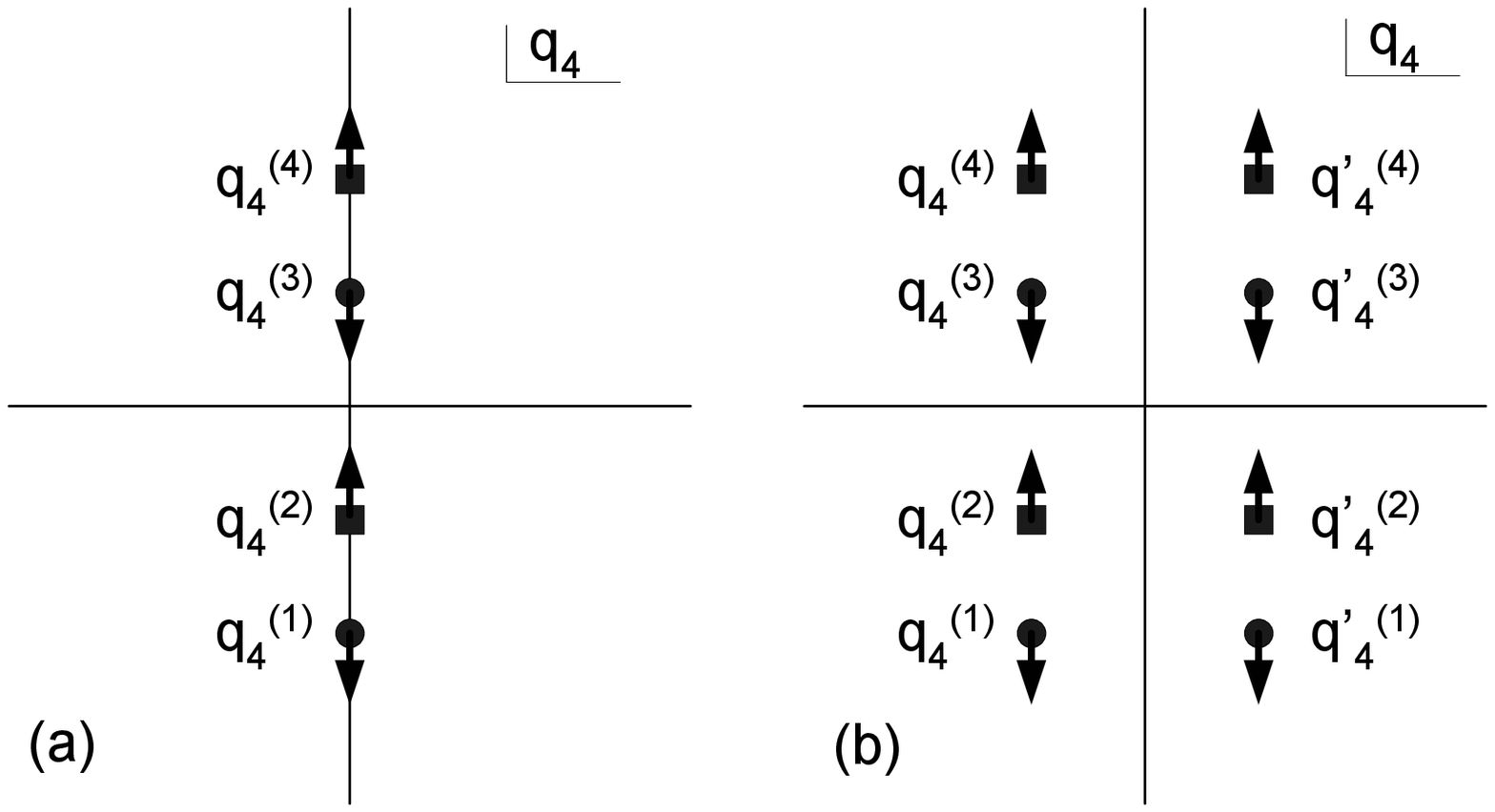,height=5cm}}
\caption{Schematic distribution in the complex $q_4$-plane of the poles
corresponding to the lowest $S_i$ set for: a) $\bar \sigma < \bar
\sigma_c$ ; b) $\bar \sigma > \bar \sigma_c$. In both cases the dots
indicate the poles of $D^+$ and the squares those of $D^-$. These
distributions correspond to a value of $p_0< 2 S_i$. The arrows indicate
the movement of the poles as $p_0$ increases. \label{figa1}}
\end{center}
\end{figure}

\begin{figure}[ht]
\begin{center}
\centerline{\psfig{figure=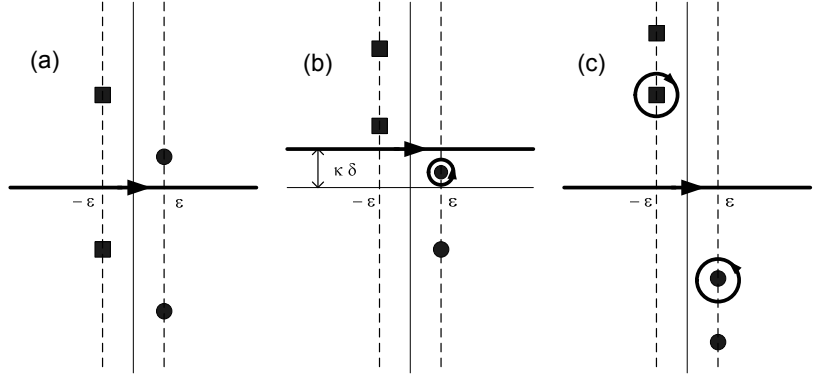,height=5cm}}
\caption{Integration paths in the complex $q_4$ plane for: a) $q_3 >
\sqrt{ \frac{p_0^2}{4} - m^2}+\delta$; b) $\sqrt{ \frac{p_0^2}{4} - m^2}-
\delta < q_3 < \sqrt{ \frac{p_0^2}{4} - m^2}+\delta$ ; c) $q_3 < \sqrt{
\frac{p_0^2}{4} - m^2}-\delta$. The constant $\kappa$ is an arbitrary real
number satisfying $\kappa > 1$. \label{figa2} }
\end{center}
\end{figure}


\begin{thebibliography}{99}
\bibitem{foot1} It can be seen that our expressions for
$g_{\pi\gamma\gamma,0}$ reduce to the result quoted in Appendix A of
Ref.~\cite{PB98}, where the steps leading to Eq.~(\ref{relgpgg}) are given
in detail.
\bibitem{foot2} Here and in what follows we use the term ``complex pole'' to
indicate a pole for which both real and imaginary parts are different from
zero.
\end{thebibliography}
\end{document}